\documentclass[12pt,preprint]{aastex}

\newcommand\de{\delta}

\newcommand\ph{\Phi}

\newcommand\<{\langle}
\renewcommand\>{\rangle}

\newcommand\beq{\begin{equation}}
\newcommand\eeq{\end{equation}}
\newcommand\bea{\begin{eqnarray}}
\newcommand\eea{\end{eqnarray}}
\newcommand\bal{\begin{align}}
\newcommand\eal{\end{align}}

\newcommand\fr{\frac}
\newcommand\tha{\theta}

\newcommand\half{{\textstyle \frac{1}{2}}}

\newcommand\cd{\cdot}

\newcommand\bk{\bold{k}}

\newcommand\br{\bold{r}}

\newcommand\by{\bold{y}}

\renewcommand\bal{\mbox{\boldmath$\alpha$}}

\newcommand\bth{\mbox{\boldmath$\theta$}}

\begin{document}

\title{On the absence of shear from perfect Einstein rings
and the stability of geometry}

\author{Richard Lieu}
\affil{Department of Physics, University of Alabama, Huntsville, AL 35899.}

\begin{abstract}

Concordance cosmology points to a Universe of zero mean curvature,
due to the inflation mechanism which occurred soon after the Big
Bang, while along a relatively small number of lower redshift light
paths where lensing events are observed, space is positively curved.
How do we know that global geometry and topology are robust rather
than in a state of chaos?  The phenomenon of cosmic shear provides
an effective way of mapping curvature fluctuations, because it
affects {\it any} light rays whether they intercept mass clumps or
not.  We discuss a range of astrophysical applications of the
principal manifestation of shear - the distortion of images.  It
will be shown that the quickest way of testing the existence of
shear in the near Universe is to look at the shape of Einstein
rings.  The fact that most of these rings are circular to a large
extent means, statistically speaking, shear occurs at a much lower
level than the expectation based upon our current understanding of
the inhomogeneous Universe.   While inflation may account for the
mean geometry, it offers no means of stabilizing it against the
fluctuations caused by non-linear matter clumping at low redshift.
Either this clumping is actually much less severe, or the physical
mechanism responsible for shaping the large scale curvature has been
active not only during the very early epochs, but also at all
subsequent times.  Might it be the vital `interface' between
expansion on Hubble distances and gravity on cluster scales and
beneath?

\end{abstract}

\keywords{ }

\section{Introduction}

In the prevailing $\Lambda CDM$ cosmological model the observational
evidence for departures from the standard theory of gravity (General
Relativity) at large distance scales, be they in the form of flat
rotation curves for galaxies or accelerated expansion between
galaxies, is explained in terms of an `extension' of the theory to
postulate two foreign ingredients: dark matter and dark energy.
Apart from the lack of direct detection of a single molecule or
quantum from either `dark' components, and especially despite
decades of large investments in the search for dark matter, the
delicate balance of proportions between the two components that
manifests itself in the observed global flatness of space
necessitates yet another postulate, viz. that the early Universe
underwent a brief period of extremely rapid expansion called the
inflation epoch.

None of the three new postulates (nor for that matter even the
expansion of space itself) have at all been verified in the everyday
laboratory, and one should also add the expansion of space to the
list, as this has recently been deemed {\it unverifiable}  by any
terrestrial apparatus (Chodorowski 2007).  Since astronomers did not
enjoy their status as pioneers of modern science through Newton's
habit of invoking unknowns to explain unknowns, it seems reasonable
to ask whether the surprises presented to us by the cosmological
data are due in fact to a breakdown in our understanding of the
nature of the fundamental forces at work over very long ranges.
Indeed, while General Relativity was comparatively well tested over
stellar distances scales, the evidence for its validity over
galactic scales and beyond are scarce and highly indirect.  It would
be very important, e.g., to be able to test if there really is the
expected statistical effect on the propagation of light by the
gravity of `embedded' mass clumps and the expansion of space in
between clumps.

Let us examine more closely this last point, as it is the subject of
the present paper.  What is the conventional understanding?  The
gravitational effect of a mass structure on light may be quantified
in terms of the impact parameter $\vec b$ between a bundle of light
rays and the center-of-mass of the distribution. If this point of
closest approach was reached during redshift $z$ and at a comoving
distance $x$ from the observer, the average fractional magnification
$\eta$ of a light source at distance $x_s > x$ away, to which the
rays look back, is given by the standard formula
\begin{eqnarray}
\eta = \frac{(x_s -x)x}{2(1+z)x_s} \left(\frac{\psi}{b} +
\frac{d\psi}{db}\right), \nonumber
\end{eqnarray}
where $\psi$ is the deflection angle.  The two terms on the right
side correspond respectively to half the fractional increase in the
linear size of the image (in angular space and relative to the image
in the absence of the mass $m$) along the directions perpendicular
and parallel to $\vec b$.

If, under what we classify here as phenomenon (a),  the light rays
pass {\it through} a mass clump, in general we will have $\eta >$ 0,
i.e. the source will be magnified. These are the lensing events.
They tend to be observationally dramatic, though their occurrence
frequency is comparatively low. If the light passes in between
clumps, two phenomena are at play: (b) the shear effect of all the
non-intercepting clumps, and (c) demagnification in the underdense
(or void) regions.  Although the effect of (c) is small, it occurs
often because the voids are large: void passages are inevitable.
Thus as a statistical average (c) cancels (a), so that the mean
angular size distance of the source is still determined only by the
overall density $\Omega$ of the Universe (Kibble \& Lieu 2005).

The scenario of particular interest to us in this paper is, however,
(b).  When our small bundle of light rays pass {\it by} (or skirts)
the clump, then $\psi= 4Gm/(c^2 b)$, so that $\psi/b$ is positive
while $d\psi/db = -\psi/b$ is negative, i.e. the resulting zero
magnification ($\eta =$ 0) is due to squeezing of the source along
the $\vec b$ direction and stretching of it in the perpendicular
direction.  This is known as shear, or weak lensing, and because in
principle any clump, no matter how far, can affect the light signal
in question, it means the shearing of a distant source is a
cumulative (hence large) effect of numerous clumps, and can occur no
matter in which direction the source lies.  Thus in (b) one is
dealing with both a significant and frequent phenomenon, which is
why weak lensing survey of many background quasars (e.g. Dodelson et
al 2006) represents a powerful technique of probing global geometry.
We shall therefore focus on the phenomenon of shear for the rest of
the paper by investigating its principal manifestation of direct
image distortion.


\section{The statistical effect of shear: how foreground galaxies can affect the appearance of distant sources}

At low redshifts any passing light can be sheared by the large scale
inhomogeneity of primordial matter {\it and} and the distribution of
non-linear virialized structures over smaller distances.  We first
examine the former.  Denote the gravitational perturbation of an
otherwise zero curvature Friedmann-Robertson-Walker Universe (as
inferred from WMAP1 and WMAP3, viz.  Bennett et all 2003 and Spergel
et al 2007) by $\Phi (x, {\bf y})$, where the $x$-axis is aligned
with the light path and ${\bf y}$ is a vector along some direction
transverse to ${\bf x}$. The correlation function between the
deflection angles $\delta {\bf y}({\bf \theta'})/x$ and $\delta {\bf
y} ({\bf \theta''})/x$ of two light rays making small angles ${\bf
\theta'}$ and ${\bf \theta''}$ w.r.t. the x-axis may be written (see
Lieu \& Mittaz 2007 for details) as
 \begin{eqnarray}
 C_{ij}(|\bth'-\bth''|)&\equiv&\fr{1}{x^2}
 \<\de y_i(\bth')\de y_j(\bth'')\>\nonumber\\
 &=&\fr{4}{c^4 x^2}\int_0^xdx'(x-x')\int_0^xdx''(x-x'')
 \<\nabla'_i\ph(x',\bth'x')\nabla''_j\ph(x'',\bth''x'')\>.
 \end{eqnarray}
We can calculate the integrals by expressing the integrand in terms
of the matter power spectrum $P(k)$,
\begin{equation}
\<\nabla'_i\ph(\br')\nabla''_j\ph(\br'')\> =
 \fr{9\Omega_m^2 H_0^4}{32\pi^3}
 \int \fr{d^3\bk}{k^3}k_ik_je^{i\bk\cd\br} P(k)
\end{equation}
where $\br = \br' - \br''$ and
\begin{equation}
P(k) = \frac{8\pi^2}{9\Omega_m^2 H_0^4}\frac{d}{d\ln
k}(\delta\Phi_k)^2,
\end{equation}
with $\delta\Phi_k$ being the standard deviation of the potential
over length scales $2\pi/k$.  The ensuing functional form of
$C(\bth) = C_{ii}(\bth) \equiv \<\de y_i(\half \bth) \de y_i(-\half
\bth)\>/x^2$ (where the repeated $i$ index implies summation over
the two ${\bf y}$ directions $(0,1,0)$ and $(0,0,1)$,  transverse to
the light path vector $(1,0,0) = \hat {\bf x}$) is
\begin{equation}
C(\bth)=C_0+\half C_2\tha^2+\mathcal{O}(\tha^4),
\end{equation}
when expanded as a Taylor series in $\theta$.

The random {\it relative} deflection between the two rays is then
given by
$$
(\delta\bth)^2 = \fr{1}{x^2}\<[\de\by(\half \bth)-\de\by(-\half
\bth)]^2\>
 =2[C(0)-C(\bth)] = C_2 \theta^2 +\mathcal{O}(\tha^4),
$$
and has the $C_2$ coefficient as its leading term, viz. $\delta
\theta \approx \sqrt{C_2} \theta \sim \theta$, clearly indicating
that the $C_0$ (constant) term relates only to {\it absolute}
deflection of the two rays.  The calculation of $C_2$, and hence
$\delta\theta$, was done in Lieu \& Mittaz (2007).  The result
points to a small shear effect, $\delta\theta/\theta = \sqrt{C_2}
\lesssim$ 1 \%, for a primordial spectrum $P(k)$ derived from
WMAP1/2dFGRS.  This is also consistent with previous conclusions
reached by Seljak (1996) and Lewis \& Challinor (2006).

Obviously, the above treatment does not take into account the role
of non-linear matter clumping; in particular galaxies in the near
Universe.  We therefore proceed to calculate the same lowest order
effect of relative deflection, i.e. $\delta\theta \sim \theta$, due
to a random ensemble of nearby galaxies.  The validity criterion of
using a Poisson clump distribution were already enumerated in detail
in section 3 of Lieu (2007); in short, the two rays must always be
separated by lengths small compared with the typical value of the
minimum impact parameter $y_{{\rm min}}$ at which each ray skirts
the galaxies.  The test for this form of shear that we shall make,
generally involves rays that satisfy this requirement.

Let us first work out carefully the effect of one mass clump.
Referring to Figure 1, the deflection angles of two neighboring
light rays with the mass clump $m$ positioned at $\by$ and $\by +
\delta\by$ relative to the points of closest approach of the rays,
are given by
 \beq
 \bal=\frac{4Gm}{c^2\by^2} \by;~~\bal' =
 \frac{4Gm}{c^2(\by + \delta\by)^2} (\by + \delta\by).
 \eeq
Denoting the angle between $\by$ and $\delta \by$ as $\vartheta$,
the differential deflection may be written as
 \beq
 \delta\bal = \frac{4Gm\delta y (\delta {\hat \by} - 2\cos
 \vartheta {\hat \by})}{c^2 \by^2}.
 \eeq
The variance in $\delta\bal$ may now be calculated.  It is
 \beq
  (\delta\bal)^2 = \left(\frac{4Gm \delta\by}{c^2
 \by^2}\right)^2,
 \eeq
and is independent of $\vartheta$.  If the rays originated from two
points that subtend the angle $\bth$ at the observer O, then Eq. (7)
will once again give the value of $\delta y$ for a mass clump at
comoving distance $x$ from O.   Moreover, we may also write
$(\delta\bth)^2 = (\delta\bal)^2$ as the variance of the random
excursion of the angular separation $\theta$ between the actual
images of these two sources.  Thus we arrive at the equation
 \beq
 (\delta\bth)^2 = \left(\frac{4Gm x\bth}{c^2 \by^2}\right)^2
 \eeq
for the shear distortion of the shape of extended sources, if $\bth$
is the angle subtented at O by two boundary points of the source.

Our final step is to derive the total variance by integrating
$(\delta\bth)^2 \times 2\pi n y~dy~dx$, where $n$ is the non-evolving
number density of clumps (for the effect of evolution see the end
of this section), down the light path $y$ and over all
impact parameters $y$ from $y=y_{{\rm min}}$ updwards.  Care should be
taken here, however, because for deflections at finite $z$ the
impact parameter scales as $y/(1+z)$ where $y$ is the comoving
distance of closest approach, which means $(\delta\bal)^2 \sim
(1+z)^2$.  If the two light rays originated from points of the same
comoving distance $D$ from O, one obtains in this way
\beq
(\delta\bth)^2 = \frac{16\pi G^2 m^2}{c^4}
\frac{n \mathfrak{D}^3  \theta^2}{y_{{\rm min}}^2},
\eeq
where $n$ is the number density of clumps, and
 \beq
 \mathfrak{D}^3 = \int_0^D  x^2 [1+z(x)]^2 dx =
 \left(\frac{c}{H_0}\right)^3 \int_0^z \frac{dz' (1+z')^2}{E(z')}
 \left[\int_0^{z'} \frac{dz''}{E(z'')}\right]^2,
 \eeq
with the function $E(z)$ being defined as
 \beq
 E(z) = [\Omega_m (1+z)^3 + \Omega_{\Lambda}]^{\frac{1}{2}}.
 \eeq
We see in Eq. (9) the phenomenon of random walk, viz.
$\delta\bth \sim \sqrt{n}$, due to the accumulation of
relative deflections along the light path from our statistical
ensemble of clumps.  There is however a divergence w.r.t.
$y_{{\rm min}}$, in the sense that the theoretical lower
limit of $y$ is zero.  In practice we conservative set
$y_{{\rm min}}$ at the value where, throughout the entire light
path, the expected number of uniformly and
randomly distributed clumps having this impact parameter is on
average equal to one.  Thus we find that for a typical direction to
some source,
 \beq
 y_{{\rm min}} = \frac{1}{\sqrt{\pi n D}}.
 \eeq
It follows that the percentage variation in the angular separation
$\theta$ between the images of the two points can be expressed as
 \beq
 \frac{\delta\theta}{\theta} = \frac{3 H_0^2}{2c^2} \Omega_{{\rm cl}}
 D^{\frac{1}{2}} \mathfrak{D}^{\frac{3}{2}},
 \eeq
after employing the definition that $Gnm \approx G \langle nm \rangle =
\sum_i Gn_i m_i$ (since there is in reality a
spread in the mass and number density of galaxies), equals
$3H_0^2\Omega_{{\rm cl}}/(8\pi)$, with $\Omega_{{\rm cl}}$ being the
{\it total} mass density of clumps as a fraction of the critical density.
The implication of this last step is that our final answer
is not too sensitive to the details of the mass function of 
the clumps, but depends primarily on the
total fraction of the mass density
$\Omega_{{\rm cl}}$ of matter belonging to all these clumps.

For application to strong lensing observations the calculation has
to be divided into two parts.  The contribution to $(\delta\bth)^2$
from galaxies in the foreground region between O and the lensing
plane is obtained from Eq. (9) with $D$ replaced by $D_l$, as
 \beq
 (\delta\bth)^2 = \frac{
 (4\pi Gm)^2}{c^4} n^2 D_l \mathfrak{D}_l^3 \theta^2.
 \eeq
where $\mathfrak{D}_l$ is as in Eq. (10) with $D_l$ (or $z_l$) as
the upper integration limit.  Next, the contribution from galaxies
lying behind the lens and in front of the source is calculated in a
likewise manner; in particular $y_{{\rm min}}$ is again from Eq.
(12) with the substitution  $D \rightarrow D_{ls}$. The total
variance is then given by
 \beq
 \frac{\delta\theta}{\theta} = \frac{3H_0^2}{2c^2} \Omega_{{\rm
 cl}} D_l^{\frac{1}{2}}\left(\mathfrak{D}_l^3 +\frac{D_l}{D_{ls}}
 \mathfrak{D}_{ls}^3 \right)^{\frac{1}{2}},
 \eeq
with
 \beq
 \mathfrak{D}_{ls}^3 = \int_{D_l}^{D_s} (D_s - x)^2 (1+z)^2 dx =
 \frac{c}{H_0} \int_{z_l}^{z_s} \frac{dz (1+z)^2}{E(z)}
 \left[D_s - \frac{c}{H_0} \int_0^{z}
 \frac{dz'}{E(z')}\right]^2,
 \eeq
when this variance is also cast as a fractional  deviation.

\clearpage
\begin{figure}
\begin{center}
\includegraphics[angle=0,width=7in]{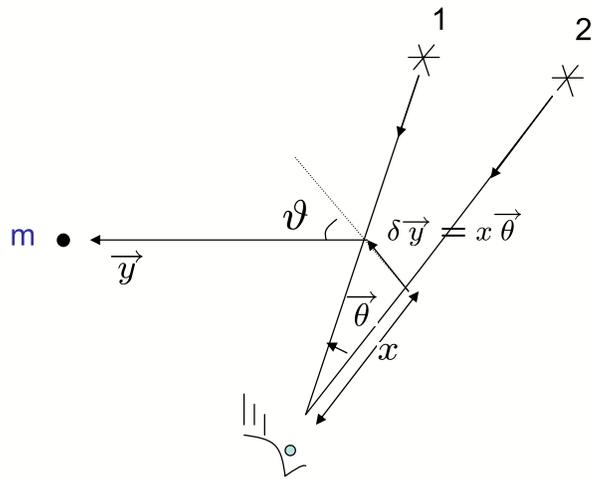}
\end{center}
\caption{Two closely spaced light rays skirting a mass
clump $m$ with impact parameters $\vec y$ and $\vec y + \delta\vec
y$, (taking into account directions).  The rays are deflected at
distance $x$ from the observer.}
\end{figure}
\clearpage

An important (and tacit) assumption underlying Eq. (15) is that
$y_{{\rm min}}$ should always remain greater than the size of a
galaxy plus its halo before we can defend our neglect of clump
evolution along the light path. Between $z=0$ and $z \sim$ 3 where
strongly lensed sources are found, evolution causes the halo of a
galaxy that is virialized at $z=$ 0 to become less compact at higher
$z$, but most of the matter in the present day halo of this galaxy
would still have `turned around' (see Eke et al 1996)
by\footnote{Thus e.g. in an Einstein-de-Sitter Universe a galaxy
just virializing today at $t=t_0$ would have turned around at
$t=t_0/2$, or $z \approx$ 0.7, when the (turnaround) radius was
$\approx$ 3.3 times larger than the $z=0$ virial radius, i.e. the
$t=t_0/2$ sphere that contains one virial mass at $t=t_0$ was a
factor of 3.3 greater in radius then, and all the matter within it
already belonged to the clump. Most galaxies that exist today would
have virialized at $z >$ 0, hence their turnaround epochs were at $z
>$ 0.7.  If dark energy is invoked to accelerate the expansion, this
would push the turnaround epoch to even higher $z$, because it would
take longer for the clump to collapse and virialize. Thus it is
reasonable to assume that when one looks back to $z \lesssim$ 3 most
galaxies were equally massive, just a few times bigger in size.} $z
\sim$ 3. Thus, the light path near the source may legitimately be
considered as being affected by the same population of random clumps
as that in $z=0$, unless the comoving scale height of the mass
distribution of the $z>$ 0 galaxies exceeds $y_{{\rm min}}$.

To be more quantitative, one may start with the observed density of
galaxies $n=$ 0.17$h^{-1} =$ 0.06 Mpc$^{-3}$ for $h =$ 0.7 (Ramella
et al 1999), to estimate that throughout the 3 Gpc comoving distance
between $z =$ 0 and $z =$ 1 a typical light ray is within $y_{{\rm
min}} \approx$ 40 kpc from a galaxy, which may be taken as an
isothermal sphere of circular velocity $\sim$ 250 km~s$^{-1}$, i.e.
the cutoff radius is then 20 kpc (this is actually a `worst case
scenario', because most galaxies are dwarfs and have radii smaller
than 20 kpc).  Even taking into account the fact that at higher $z$
a virialized system was effectively larger by the factor $1+z$, at
$z =$ 1 the average comoving galaxy radius then becomes 40 kpc,
which is barely equal to $y_{{\rm min}}$.  Thus as mentioned before
the passing light generally misses all the galaxies.  If, however,
evolution causes the galactic halo to lie beyond the 20$(1+z)$ kpc
virial radius in the past, so that $y_{{\rm min}}$ falls within the
halo scale height back then, the light rays would have been weakly
lensed at that part of their journey, and all shear effects will be
reduced from the level calculated above because the mass $m$ that
affected the light falls short of the galaxy's total mass.  Such a
violation of our present assumption could take place on the far side
of the strong lensing plane where $z \gtrsim$ 1, thereby lowering
$\delta\theta/\theta$ to a value given only by the first term of Eq.
(15).  On the near side of the strong lensing plane evolution is
less important because, galaxies generally virialize well ahead of
the $z =$ 0 epoch. This theoretical (modeling) result is also
corroborated by observations, which indicate that galaxies indeed
exhibit no evidence for evolution at least up to redshifts $z
\approx$ 1 (Ofek et al 2003).

We close this section with a point of fundamental physics.  Whether
the cause be primordial matter or non-linear clumping, the relative
deflection between two neighboring rays as calculated above
originates from the first order ($C_2$) term in the Taylor expansion
of Eq. (16), and is the same order effect as the incoherent time
delay of Lieu (2007).  In fact, by means of the light reciprocity
theorem it was shown (Lieu \& Mittaz 2007) that the formulae for
relative deflection and incoherent delay are inter-convertible, and
that the same statement applies to absolute deflection and coherent
delay. Although the latter pair are both zeroth order effects they
are much harder to observe, as already explained in Lieu (2007).

\section{The shape of distant galaxies; superluminal motion in quasars}

Here we discuss two astrophysical applications of section 2 in the
context of $\Lambda$CDM cosmology, where
 \beq
 \Omega_m = 0.3,~\Omega_{\Lambda} = 0.7,~h=0.7,~{\rm
 and}~\Omega_{{\rm cl}} = 0.15.
 \eeq
The setting of $\Omega_{{\rm cl}} =$ 0.15 is fully consistent with
the expectation of the standard model, which assumes that half the
baryons, hence approximately the same fraction for dark matter also,
of the low $z$ Universe resides in galaxies and their halos - mass
clumps that may completely be distributed as field galaxies or
partly congregated into groups, see Fukugita (2004) and Fukugita et
al (1998). In fact, the observed properties of galaxies given in the
last section do indeed yield $\Omega_{{\rm cl}} =$ 0.15.

In the first application we consider the appearance of resolved
sources at $z \gtrsim$ 1.  According to Eqs. (15) and (17)
$\delta\theta/\theta \geq$ 13 \% at $z \geq$ 1.  Since this is
caused by shear, viz. the light rays in question have not directly
been lensed, there is no magnification (section 1), Hence, if the
percentage change in the angular size of the image along one
direction has the typical value of +13 \%, the same for the
orthogonal direction must be -13 \% (i.e. between the two axes
$\delta\theta/\theta$ are correlated) to conserve total solid angle
subtended by the image at us.  This means the aspect ratio of the
resulting distorted image reaches 26 \% at $z =$ 1, and larger at $z
>$ 1. Now most $z \gtrsim$ 1 sources we detect are quasars, i.e.
elliptical galaxies to `begin with'.  One could ask if the
ellipticity is completely due to shear.  Nevertheless, an aspect
ratio $\gtrsim$ 26 \% is quite large, so that even ahead of a
statistical analysis at high resolution one could already query if
such a level of shear really exists.  Thus, apart from time delay
observations, this represents another potential challenge to all the
cosmological models, with $\Lambda$CDM in particular.

Next we turn to the problem of quasar superluminal motion, which is
inferred from the angular speed at which blobs of ejected material
move away from the central AGN engine: by means of the distance to
the quasar as derived from its redshift, this angular speed is often
converted to a physical speed that exceeds $c$.  When the redshift
of the quasar is high, however, caution is needed in the conversion,
because if in Eq. (15) $\delta\theta/\theta$ is no longer $\ll 1$,
the true angular speed could be substantially larger or smaller than
the observed value, depending on which way the apparent motion of
the blob is being sheared by the foreground matter.  In fact,
$\delta\theta/\theta$ is the percentage error in the superluminal
speed (note that because superluminal motion typically involves
$\theta \sim$ milli-arcseconds, i.e. the light rays being sheared
are very close to each other, the assumptions underlying the
validity of Eq. (15) holds exceptionally well).  Thus, the point
raised here starts to be relevant for quasars of $z \gtrsim$ 1,
where $\delta\theta/\theta \gtrsim$ 13 \% as before.

As the improvement of sensitivity and resolution may lead to the
discovery of quasar superluminal motion at higher redshifts and
(likely) with ever increasing jet speeds, examples being Bouchy et
al 1998 on 1338+381 at ($z =$ 3.1, $v_{\perp}/c \approx 27/h$) and
Frey et al 2002 on 1351-018 at ($z=$ 3.7, $v_{\perp}/c \approx
9.2/h$), it is {\it an expectation, based upon the cosmological
effect of shear, that not all of the apparent largeness of
$v_{\perp}/c$ is due to relativistic distortions at the source.}  An
interesting future pursuit worthy of consideration is to correlate
$v_{\perp}/c$ with $z$ to see if there is more scatter at high
redshifts.  If so, this could be indicative of the presence of
shear.

\section{Why are perfect Einstein rings a challenge to cosmological models?}

Finally, the third application of section 3.  We return to the
question raised in section 1, on whether existing data can already
be used to {\it clinch} cosmological models on the problem of global
geometry.  We hold the view that the most effective test {\it
currently} available is still the weak lensing distortion of images
of distant sources mentioned in section 1. The new point to be made
in this work, however, is that unlike the primordial matter
distribution the shear effect of foreground galaxies is severe for
$z \gtrsim$ 1 sources, i.e. one should not need such a large sample
of background emitters to detect it. Nevertheless, the usual
difficulty is in finding circularly symmetric patterns  to start
with, so that one knows that any apparent elongation is not an
intrinsic property of the object being looked at.

For the above reasons the Einstein rings of well-aligned strong
lensing configurations, play a unique role in satisfying our
requirement, because the intrinsic shape of such a pattern is
circular, or quite nearly so.  While the zeroth order `$C_0$' term,
or absolute deflection at constant $\delta\theta =$ by mass
inhomogeneities can affect the {\it existence} of an Einstein ring
by bringing misaligned source-lens-observer arrangements into
alignment (i.e. in a smooth Universe the same Einstein ring seen
somewhere in the sky would not even have been observable as the
optical components involved are intrinsically non-collinear), the
higher order effect of relative deflection ($\delta\theta \sim
\theta$) between two neighboring rays calculated in section 2 plays
the role of {\it distorting} the ring via cosmic shear, as
illustrated in Figure 2.

We therefore focus our attention upon a very recent set of well
observed Einstein ring images, starting with the best candidate,
J0332-3357, where $z_l =$ 0.986, $z_s =$ 3.773 (Cabanac et al 2005),
or $D_l =$ 3.271 Gpc,  $D_s =$ 7.012 Gpc, $D_{ls} = D_s - D_l$ in
the flat $\Lambda$CDM cosmology of Eq. (17).  Substituting these
numbers into Eq. (15), one obtains $\delta\theta/\theta =$ 25 \% or
aspect ratio 50 \%.  It is evident without any further analysis
necessary that the J0332-3357 Einstein ring is much too circular to
accomodate such a significant ellipticity.

Could the J0332-3357 observation simply be a statistical anomaly?
There has recently been a wave\footnote{See the images on
\texttt{http://hubblesite.org/newscenter/archive/releases/2005/32/image/a/}.}
of Einstein ring detections, such as J073728.45+321618.5,
J232120.93-093910.2, and J163028.15+452036.2. Together with the more
historical B1938+666 (King et al 1997), all these lensing systems
have characteristic parameters values $D_l \approx D_{ls} \approx$
3.3 Gpc ($z_l \approx$ 1, $z_s \approx$ in the cosmology of Eq.
(17)), i.e. the resulting $\delta\theta/\theta$ of shear is at a
comparable level as that for J0332-3357, and can be cast in a
convenient form as
 \beq
 \frac{\delta\theta}{\theta} = 0.23 \left(\frac{h}{0.7}\right)^2
 \left(\frac{\Omega_{{\rm cl}}}{0.15}\right)
 \left[\left(\frac{D_l}{3.3~{\rm Gpc}}\right)^{\frac{1}{2}}
 \left(\frac{\mathfrak{D}_l^3}{35.7~{\rm Gpc}^3}\right) +
 \left(\frac{D_l}{D_{ls}}\right) \left(\frac{\mathfrak{D}_{ls}^3}
 {71.3~{\rm Gpc}^3}\right)\right]^{\frac{1}{2}}
 \eeq
(noting that the $D$'s and $\mathfrak{D}$'s are related to each
other once a cosmology is chosen).  Thus these images should also be
sheared with an aspect ratio similar to that of J0332-3357, yet none
of them are observed to exhibit this behavior. As a guide to the
eyes, we show in Figure 3 an ellipse with 50 \% aspect ratio. It is
fair to say that no continuous and nearly-complete Einstein rings
have been found to suffer from so much distortion.

Since our prediction on shear, Eq. (15), lies with the fact that it
depends simply on the mean mass density of clumped matter
$\Omega_{{\rm cl}}$ (which cannot differ too greatly from
$\Omega_{{\rm cl}} \approx \Omega_m/2$ or else structure formation
will be in jeopardy, see sections 2 and 3) regardless of e.g. the
number density of clumps or the mass distribution of individual
clumps, this goes to highlight just how robust the prediction is.
Nevertheless,  Eq. (15) has caveats and these were stated in section
2, where we contended that the details on clumping are irrelevant
provided the Universe did not have too high a degree of homogeneity
at any time during the light propagation.  Specifically a milder
shear prediction can be reached by appealing to the increased size
of galaxies and groups at higher redshifts as the Universe turned
too smooth (section 2).  In a revised prediction which is probably
over conservative, one could take into account the shear
contribution, in the manner calculated in this and the last section,
only from those clumps lying within the foreground Universe between
the (strong) lensing plane at $z \approx$ 1 and the observer, i.e.
considering the Universe behind the lensing plane as completely
homogeneous. According to section 3, the expected aspect ratio of an
Einstein ring would then be $\sim$ 26 \%.  Such a distortion is also
depicted in Figure 3, to demonstrate that it is still too large to
reflect reality - among the aforementioned Einstein rings only
J140228.21+632133.5 exhibit a commensurate ellipticity.

\vspace{3mm}

\section{Conclusion: the question of global geometry}

When light propagates through an inhomogeneous Universe there are
three possible outcomes: (a) if the rays intercept mass clumps there
will be magnification, (b) if they only skirt a clump there there
will be shear, (c) if they intercept underdense voids there will be
demagnification.  When the mean density is critical, the effects of
(a) and (c) cancel statistically, and in this sense one can say that
the `mean' global geometry of space is determined solely by
$\Omega$. In terms of occurence rate, however, class (a) events are
relatively rare, and although (c) are more frequent each event
brings about a small change which is hard to measure.  On the other
hand, for any bundle of light rays(b) applies to {\it all} clumps
with impact parameters less than a Hubble radius, and the sum total
of their contributions is a rather large shear which has observable
consequences, such as the appearance of high redshift quasars and
systems with superluminal motion.

But perhaps most important manifestation of shear is to be found in
the apparent shape of the (intrinsically circular) Einstein rings,
as these should be stretched into randomly oriented ellipses with an
aspect ratio $\gtrsim$ 25 \%.  Observed rings are, however, usually
much less affected, so unless the near Universe is much more
homogeneous than expected this poses a formidable challenge to
cosmological models, viz. if geometry is shaped solely by inflation
rather than some physical mechanism that operates at all epochs, why
such fluctuation as shear are not seen in the Einstein rings?

It is therefore not entirely inconceivable that the puzzles of
cosmology in general and global geometry in particular lie
hand-in-hand. If the latter is solved, we will be much closer to
developing the correct model of the former. Current distinction
between the concepts of `dark energy' and `dark matter' merely
represents our on-going effort in seeking a complete understanding
of the missing physical mechanism(s) that bridge the gap between
gravity on smaller scales, and expansion of space on Hubble scales.
Though at present an improbable scenario, the stability of a certain
prescribed geometry might have been enforced by an ongoing tension
between two opposing mechanisms, i.e. the gravitational fields of
mass clumps are not simply `embedded' phenomena in an otherwise
decoupled `ether' of uniform and accelerated expansion.

The author is grateful to Prof. S. -N. Zhang at Tsinghua University,
Beijing, for his hospitality during a very pleasant two-month visit in the
summer of 2007 when most of the thoughts in this paper were conceived
and written.  He 
is also indebted to Zhang's PhD student
Ally Jiang for her preparation of
all three figures.

\clearpage
\begin{figure}
\begin{center}
\includegraphics[angle=0,width=.6\textwidth]{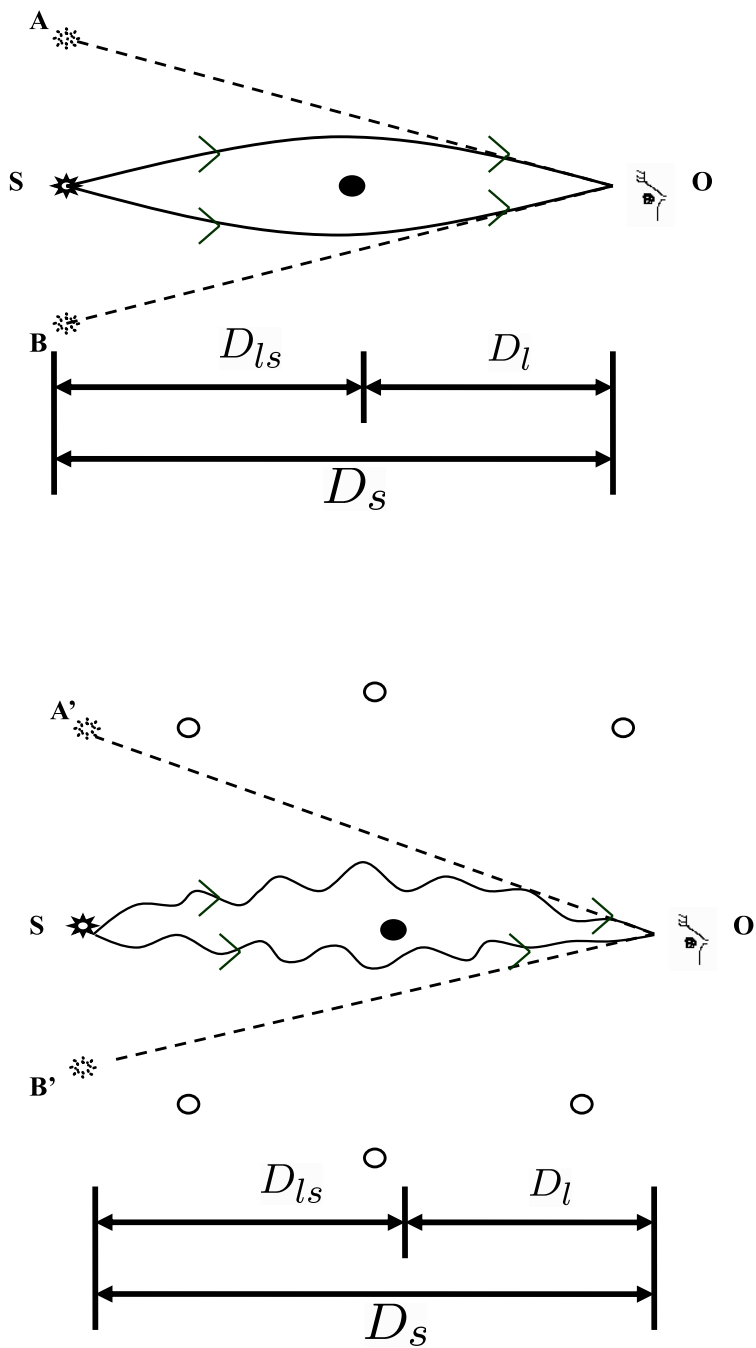}
\end{center}
\caption{Two strong lensing images (which could be part
of an Einstein ring) are displaced {\it relatively} to each other as
their associated light paths are perturbed by external mass clumps
lying at distances far larger than the separation between the paths.
This relative displacement is also accompanied by an incoherent
(random) difference in the arrival times of two photons emitted
simultaneously and propagating down the two paths.  In fact, both
the relative deflection and incoherent delay are first order
effects.  Moreover, they are manifestations of the {\it same}
underlying phenomenon, viz. shear.}
\end{figure}

\begin{figure}
\begin{center}
\includegraphics[angle=0,width=7in]{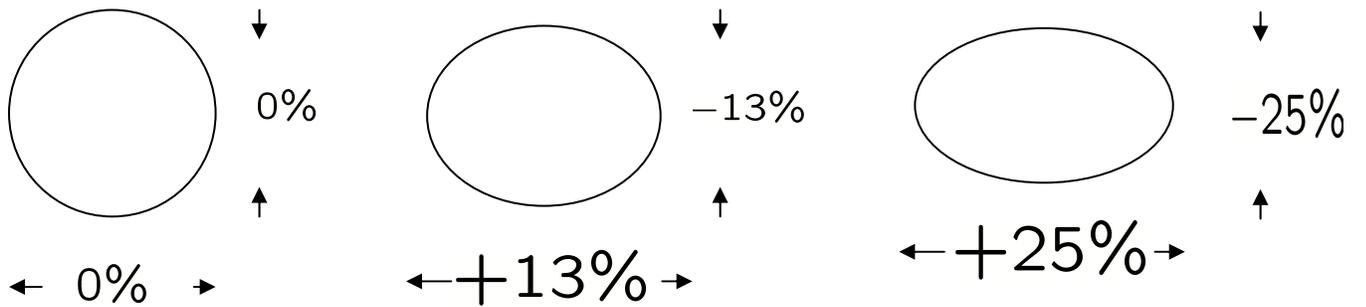}
\end{center}
\caption{The distortion of a perfect Einstein ring
(leftmost circle) by shear.  Clumps lying in the foreground, i.e.
between us and the lensing plane, could deform the ring's appearance
to become like the first ellipse.  If background clumps are also
taken into account and their evolution is neglected, the ring will
be stretched into the shape of the second ellipse.  In reality the
orientation of the two ellipses is, of course, random.}
\end{figure}

\end{document}